\documentstyle[12pt,amssymb]{article}
\normalbaselines
\catcode `\@=11
\@addtoreset{equation}{section}

\def\section{\@startsection {section}{1}{\z@}{-3.5ex plus -1ex minus
     -.2ex}{2.3ex plus .2ex}{\normalsize\bf}}
\def\subsection{\@startsection{subsection}{2}{\z@}{-3.25ex plus -1ex minus
 -.2ex}{1.5ex plus .2ex}{\normalsize\bf}}
\def\thebibliography#1{\section*{References\markboth
  {REFERENCES}{REFERENCES}}\list
  {[\arabic{enumi}]}{\settowidth\labelwidth{[#1]}\leftmargin\labelwidth
  \advance\leftmargin\labelsep
 \usecounter{enumi}}
  \def\newblock{\hskip .11em plus .33em minus -.07em}
  \sloppy
  \sfcode`\.=1000\relax}
 
\catcode `\@=12
\begin{document}

\vspace*{2cm}
\begin{center}
%\noindent
{\bf A REMARK ON THE BOSON-FERMION CORRESPONDENCE}
\vspace{1.3cm}\\
\end{center}
%\noindent
\hspace*{1.5cm}
\begin{minipage}{13cm}
Yurii A. Neretin
\footnotemark
\vspace{0.3cm}\\
  Chair of Mathematical Analysis, 
  Moscow Inst. of Electronics and Mathematics
  Bol'shoi Triohsviatitel'skii per. 3/12,  Moscow 109082, Russia,\\
  E-mail: chuhloma@neretin.mccme.ru
\end{minipage}
\footnotetext{supported by RFBR grant 98-01-00303
and by Russian program of support of scientific schools}

\vspace*{0.5cm}

\begin{abstract}
\noindent
We introduce the space of skew-symmetric functions depending
on an infinite number of variables and give a simple interpretation
of the boson-fermion correspondence.
\end{abstract}

\def\F{{\rm\bf F}}
\def\L{{\bf \Lambda}}
\def\S{{\rm\bf Symm}}
\def\A{{\rm\bf Asymm}}

\vspace{22pt}

The boson-fermion correspondence
 (Skyrme, 1971) is a canonical transformation from
bosonic Fock space to fermionic Fock space (more precisely it is
an operator from some special bosonic Fock space to some special fermionic
Fock space). Now it is a quite well-known
 object in mathematics
and mathematical physics (see for instance \cite{PS}).
The purpose of this note is to give a
very simple description of this operator.
The boson-fermion correspondence will be  multiplication with the
Vandermonde determinant. In some sense our description  is not new
(it is equivalent to an explanation which uses Schur functions, see \cite{PS}),
on the other side I never have seen this description in the literature
and never have heard about it.

{\bf 1. Bosonic Fock space $\F$.}
Consider formal variables $z_1,z_2,\dots$. Consider the space $Pol$ of
polynomials in the variables $z_1,z_2,\dots$.
Define a scalar product in $Pol$ by the following rule: the monomials
$z_1^{k_1}z_2^{k_2}\dots$
are pairwise orthogonal and
\begin{equation}
\|z_1^{k_1}z_2^{k_2}\dots\|^2= \prod_j\left(k_j! j^{k_j}\right)\ .
\end{equation}
We define the bosonic Fock space $\F$ (V.A.Fock, 1929, see \cite{PS, Ner})
 as the completion of $Pol$
with respect to this scalar product.

{\bf 2. Space $\S$ of symmetric functions.}
Consider an infinite collection of formal variables $x_1,x_2,\dots$.
We define the space $\S$
 of symmetric functions as the space of symmetric
  infinite formal sums of monomials
in the variables $x_1,x_2,\dots$ (see \cite{Mac}) (in each monomial 
only finite number of variables occurs).

 Denote by   $p_k$ the infinite Newton sums
$$p_n= x_1^{n}+x_2^{n} +x_3^n+\dots$$
The classical scalar product (J.H. Redfield, 1927) in the space $\S$ is given
by the rule: the "functions"
$p_1^{k_1}p_2^{k_2}\dots$
are orthogonal and
\begin{equation}
\|p_1^{k_1}p_2^{k_2}\dots\|^2= \prod_j\left(k_j! j^{k_j}\right)\ .
\end{equation}

{\bf 3. Boson--symmetric correspondence}, see \cite{PS}. 
A canonical
isometry $I:\F\rightarrow\S$ is  given by the rule
$$I:z_1^{k_1}z_2^{k_2}\dots\mapsto
p_1^{k_1}p_2^{k_2}\dots$$
In other words the operator $I$ is a substitution  operator
$$I\,f(x_1,x_2,x_3,\dots  )=f(\sum_j x_j,\sum_j x^2_j,\sum_jx_j^3,\dots)\ .$$

Obviously $I$ is an isometry (see (1) and (2)).

{\bf 4. Space of skew-symmetric functions.} This object is very simple
but psychologically strange. Consider the same variables $x_1,x_2,\dots$.
A  {\it quasi-monomial}  is a formal expression
$$x_1^{\omega+l_1}x_2^{\omega+l_2}x_3^{\omega+l_3}\dots$$
where $l_j=-j$ for large $j$ and $\omega$ is a formal symbol.
A {\it skew-symmetric function} is a
formal (infinite) linear combination of quasi-monomials which
is skew-symmetric with respect to all finite permutations of the 
variables $x_1,x_2,\dots$.

{\sc Remark.} Informally, $\omega$ means
           $$\omega=\infty\ .$$
It is "the total number" of variables
$x_1,x_2,\dots$. It is natural to consider the expression
\begin{equation}
\prod_{1\le i<j<\infty} (x_i-x_j)
\end{equation}
as skew-symmetric function. Indeed let us write
this expression in the form
$$\prod_{1\le i<j<\infty} \left\{x_i(1-\frac {x_j}{ x_i})\right\}
=\prod_{1\le i<j<\infty} x_i
\prod_{1\le i<j<\infty} (1-\frac {x_j}{ x_i})
                       $$

We obtain
\begin{equation}
\prod_{1\le i <j<\infty} (x_i-x_j)=\sum_{\sigma \in S_\infty}
(-1)^\sigma x_1^{\omega-\sigma(1)}x_2^{\omega-\sigma(2)}\dots
\end{equation}
where $S_\infty$ is the group of all finite permutations of
the set $\{1,2,3,4,\dots\}$.
\medskip

Let $\l_1<l_2<l_3<\dots$ be integers and let $l_j=j $ for large $j$.
Consider the basic skew-symmetric functions
$$S_{l_1,l_2,\dots}=\sum_{\sigma\in S_\infty} (-1)^\sigma
x_1^{\omega-l_{\sigma(1)}}x_2^{\omega-l_{\sigma(2)}}x_3^{\omega-l_\sigma(3)}$$
A scalar product in the space $\A$ of skew-symmetric
 functions is defined
by the rule: the functions $S_{l_1,l_2,\dots}$ form an orthonormal basis in
$\A$.

{\bf 5. Correspondence between $\S$ and $\A$.} A canonical isometry
$J: \S\rightarrow\A$ is given by the formula
$$Jf(x_1,x_2,\dots)=f(x_1,x_2,\dots)\cdot\prod_{1\le i<j<\infty}(x_i-x_j)\ .$$

{\bf 6. Fermionic Fock space}, see \cite{PS, Ner}. Let
 $\dots\xi_{-2},\xi_{-1},\xi_0,\xi_1,\xi_2,\dots$
be a family of anticommuting variables
($\xi_i\xi_j=-\xi_j\xi_i$). Consider infinite products
\begin{equation}
\xi_{l_1}\xi_{l_2}\xi_{l_3}\dots
\end{equation}
where $l_1<l_2<\dots$ and $l_j=j$ for large $j$.
We define the {\it fermionic Fock space} $\L$ as the 
space where the monomials (5) form
an orthonormal basis.

{\bf 7. Isometry between $\S$ and $\L$.} This correspondence is
obvious:  the      basis element (4) corresponds to the basis element (5).

{\bf 8. The boson-fermion  correspondence} 
is the composition of the correspondences
$$\F\rightarrow\S \rightarrow\A\rightarrow\L\ .$$
In fact it is the composition of the substitution
$$z_k=\sum_j x_j^k$$
  and the multiplication with the  Vandermonde determinant (3).

Some additionsl discussion of boson-symmetric correspondences is contained in
\cite{FA}, \cite{MS}

     \end{document}